\documentclass[preprint,groupedaddress,preprintnumbers,showpacs,amsmath,amssymb]{revtex4}

\usepackage{amsmath}
\usepackage{amssymb}
\usepackage{amsfonts}
\usepackage{mathrsfs}
\usepackage{epsfig}
\usepackage{bm}

\begin{document}

\title{Polarized Beam Conditioning in Plasma Based Acceleration}

\author{J. Vieira$^1$}
\author{C.-K. Huang$^2$}
\author{W.B. Mori$^3$}
\author{L. O. Silva$^1$}

\affiliation{$^1$GoLP/Instituto de Plasmas e Fus\~{a}o Nuclear-Laborat\'orio Associado,  Instituto Superior T\'{e}cnico, 1049-001 Lisboa, Portugal}
\affiliation{$^2$Los Alamos National Laboratory, Los Alamos, New Mexico 87545}
\affiliation{$^3$Department of Physics and Astronomy, UCLA, Los Angeles, California 90095, USA}

\today

\begin{abstract}
The acceleration of polarized electron beams in the blowout regime of plasma-based acceleration is explored. An analytical model for the spin precession of single beam electrons, and depolarization rates of zero emittance electron beams, is derived. The role of finite emittance is examined numerically by solving the equations for the spin precession with a spin tracking algorithm. The analytical model is in very good agreement with the results from 3D particle-in-cell simulations in the limits of validity of our theory. Our work shows that the beam depolarization is lower for high-energy accelerator stages, and that under the appropriate conditions, the depolarization associated with the acceleration of 100-500 GeV electrons can be kept below $0.1-0.2\%$.
\end{abstract}


\maketitle


\section{Introduction}
\label{sec:introduction}


Polarized particle beams are widely used in state-of-the-art high-energy physics (HEP) experiments \cite{bib:mane_jpg_2005,bib:mane_rpp_2005,bib:pick_pr_2008}. The beam polarization, which corresponds to the average of the particle beam spins, is an essential beam feature, critical to several fundamental physical problems, including precision tests to the standard model~\cite{bib:pick_pr_2008}, the search for the Higgs boson~\cite{bib:higgs_prl_1964}, and the discovery of new physics, for instance, in higher dimensions~\cite{bib:pick_pr_2008}. State-of-the-art and future accelerators are aiming at the acceleration of highly polarized electron beams. For instance, design for the International Linear Collider (ILC) considers 0.5 TeV electron beams with polarizations above $80~\%$ (close to the maximum polarization of $100 \%$, where all the beam particles spins are aligned in the same direction), with typical depolarizations below $0.1-0.2~\%$~\cite{bib:pick_epac_2006}, which are mainly associated with the beam-beam interaction at the interaction point.


The maximum acceleration gradients that can be attained in standard accelerators, and that directly impacts on their size, and cost, is limited by material breakdown thresholds. The use of plasma waves to accelerate particles can, therefore, play an important role in future generations of accelerators by sustaining acceleration gradients which can be more than three orders of magnitude higher than standard accelerators~\cite{bib:tajima_prl_1979}.


In plasma based acceleration (PBA), accelerating structures can be excited by either particle or laser drivers. The most striking experiments using electron beam drivers [plasma wakefield acceleration (PWFA)] doubled the energy of 42 GeV electron beams in 80 cm long plasmas~\cite{bib:blumenfeld_nat_2007}. These results were obtained in the blowout regime, where plasma electrons are evacuated from the region where the driver propagates. The resulting wakefield structures are characterized by linear accelerating and focusing forces, and are ideally suited for electron acceleration~\cite{bib:lu_prl_2006,bib:lu_prstab_2007,bib:tzoufras_prl_2008}. 

The most important advances using laser pulse drivers [laser wakefield acceleration (LWFA)], also occurred in the blowout regime, and demonstrated the acceleration of 1 GeV electron beams in cm scale plasmas~\cite{bib:mangles_nature_2004,bib:faure_nature_2004,bib:geddes_nature_2004,bib:leemans_natphys_2006,bib:kneip_prl_2009}. Several techniques were also devised in order to control the acceleration processes~\cite{bib:kalmykov_prl_2010}, adjusting the bunch energy, charge~\cite{bib:faure_nature_2006,bib:geddes_prl_2009,bib:davoine_prl_2009,bib:pak_prl_2010}, and transverse features~\cite{bib:vieira_prl_2011}.  

Recently, the use of proton drivers was also proposed as a means to accelerate electron beams even further to $\gtrsim 0.5~\mathrm{TeV}$ in $\lesssim 1~\mathrm{km}$ plasmas~\cite{bib:caldwell_nat_2009}. These results show, clearly, the capability of PBA, and in particular, the capability of the blowout regime of PBA, to accelerate high-quality electron beams to high energies~\cite{bib:martins_naturephys_2010,bib:martins_pop_2010} in controlled acceleration scenarios. However, for a future plasma-based linear collider, or for a plasma afterburner~\cite{bib:blumenfeld_nat_2007,bib:lee_pre_2000,bib:huang_pac_2009,bib:schroeder_prstab_2010,bib:seryi_pac_2009}, it is important not only to achieve high final output energies, but also to achieve low and controlled depolarization rates.


The depolarization of an electron beam is fully determined by the temporal evolution of the direction of the spin, $\mathbf{s}$, of each beam electron. During its acceleration, the spin of the electron can change according to two different mechanisms: the stochastic spin diffusion from photon emission~\cite{bib:sokolov_sp_1969,bib:shatunov_aip_2001}, and the spin precession around the electric and the magnetic fields~\cite{bib:mane_rpp_2005,bib:bargman_prl_1959}. On one hand, the stochastic spin diffusion is a non-deterministic process consisting in the rotation of the spin vector in the presence of a magnetic field, with the emission of a photon. The Sokolov-Ternov effect~\cite{bib:sokolov_sp_1969} (spin-flip) is a well-known stochastic spin diffusion mechanism, which can polarize electron beams in circular accelerators. On the other hand, the spin precession is a deterministic process, and can be examined by treating the spin as an intrinsic electron magnetic moment.

The spin precession is an important mechanism that has a decisive role in the design of accelerators. In circular accelerators, for instance, the spin precession can depolarize the beam completely if the spin precession frequency is a multiple of the orbital frequency (depolarizing resonance)~\cite{bib:mane_jpg_2005,bib:mane_rpp_2005}. In linear accelerators (linacs) and colliders, however, the typical depolarizations are as low as $0.1-0.5\%$ for $0.5-1~\mathrm{TeV}$ accelerators and colliders~\cite{bib:pick_epac_2006}. This work is then focused on the dynamics of spin precession in plasma wakes, while the nondeterministic spin diffusion is left for a future work.


This Paper shows that the electron beam depolarization during the acceleration in PBAs may fulfill the requirements for high energy physics experiments, provided that the beam emittances are sufficiently low, and shows that depolarizations of $0.1-0.2\%$ can be achieved in 100-500 GeV accelerators. In Section~\ref{sec:precession}, the Thomas--Bargman-Michel-Telegdi (T-BMT) equations, which describe the spin precession of relativistic charged particles, are used to derive a set of coupled equations for the spin precession in plasma accelerators. This analysis reveals that the spin precession is fully determined by the transverse forces that act on relativistic particles. In Sec.~\ref{sec:polarization}, analytical expressions for the depolarization associated with zero-emittance electron beams are derived. In Section~\ref{sec:sgvseg} it is shown that externally guided propagation regimes lead to lower depolarization rates in comparison to self-guided propagation regimes. In addition, it is found that the depolarization of higher energy plasma accelerators is lower than that of lower energy plasma accelerators. In Sec.~\ref{sec:simulations} the model is compared with numerical simulations in scenarios that go beyond the validity limits of the analytical theory. Two approaches were used: in the first approach, the prescribed electromagnetic fields and trajectories of electrons in the blowout regime are considered; the second approach uses the fully self-consistent electromagnetic fields and electron trajectories from 3D PIC simulations in QuickPIC~\cite{bib:huang_jcp_2006}. Finally, in Sec.~\ref{sec:conclusions}, the conclusions are stated.


\section{Single electron spin dynamics}
\label{sec:precession}

In order to determine the polarization of the beam, the spin precession of a single electron is firstly examined. For non-relativistic electrons, the spin precession follows: 

\begin{equation}
\label{eq:precession}
\frac{d \mathbf{s}}{d t} = \mu \times \mathbf{B} = -g\frac{\mathbf{s}\times \mathbf{B}}{2} ,
\end{equation}
where $\mu=-g~\mathbf{s}/2$  is the electron intrinsic magnetic moment, $g\simeq2.0023228$ is the dimensionless magnetic moment of the electron (g-factor), $t$ is the time normalized to the inverse of the plasma frequency $\omega_p=\sqrt{4\pi e^2 n_0/m_e}$, with $e$ and $m_e$ being the electron charge and mass, $n_0$ the plasma density, $\mathbf{B}$ is the magnetic field normalized to $m_e c \omega_p/e$, and where $c$ is the speed of the light. 

Eq.~(\ref{eq:precession}) indicates that the spin of a non-relativistic electron -- with relativistic gamma factor $\gamma = 1/\sqrt{1-\mathbf{v}^2} \gtrsim 1$, with $\mathbf{v}$ denoting its velocity normalized to $c$ -- precesses around the magnetic field lines with frequency $\mathbf{\Omega}^{\mathrm{NR}}=-\mu \mathbf{B} = g \mathbf{B}/2$. However, most interesting particle acceleration scenarios use ultra-relativistic electron beams with $\gamma\gg 1$. In order to obtain a correct description of the spin precession dynamics for this case, the relativistic generalization of Eq.~(\ref{eq:precession}), which is not covariant, is required, and given by~\cite{bib:bargman_prl_1959}:

\begin{equation}
\label{eq:tbmt}
\frac{d \mathbf{s}}{ d t} =-\left[\left(a+\frac{1}{\gamma}\right)\left(\mathbf{B}-\mathbf{v}\times\mathbf{E}\right)-\mathbf{v}\frac{a \gamma}{\gamma+1} \mathbf{v} \cdot \mathbf{B}  \right] \times \mathbf{s}= \mathbf{\Omega} \times \mathbf{s}.
\end{equation}
All the quantities in Eq.~(\ref{eq:tbmt}) are in the laboratory frame, except for $\mathbf{s}$, which is an intrinsic property of the electron and is, for that reason, described in the electron rest frame. Moreover, the precession frequency $\Omega$ is $\mathbf{\Omega}=[(a+1/\gamma)(\mathbf{B}-\mathbf{v}\times\mathbf{E}-\mathbf{v}a\gamma \mathbf{v}\cdot\mathbf{B}/(\gamma+1)]$. The quantity $a=(g-2)/2\simeq 0.0011614$ is the anomalous magnetic moment of the electron, and $\mathbf{E}$ is the electric field normalized to $m_e \omega_p/e$.

Although the T-BMT equations are strictly valid when $\mathbf{E}$ and $\mathbf{B}$ are homogenous, they can still be used as long as Stern-Gerlach type forces($F^{\mathrm{S-G}}$) can be neglected~\cite{bib:mane_rpp_2005}, i.e. as long as the spin dynamics does not change the electron trajectories. These forces are proportional to $F^{\mathrm{S-G}}\propto \nabla(\mu\cdot\mathbf{B})\propto (g \hbar/2) r_0 \omega_{\beta} \mathbf{B}$, where $\mathbf{B}\propto r_0 \cos(\omega_{\beta} t)$ was assumed, and where $\omega_{\beta} = \omega_p/\sqrt{2 \gamma}$ is the betatron frequency. Thus, Lorentz forces, proportional to $F^{\mathrm{L}}\propto\mathbf{E}+\mathbf{v}\times \mathbf{B} \propto r_0 \gg F^{\mathrm{S-G}} \propto \hbar r_0^2 /\sqrt{\gamma}$, dominate the electron dynamics in the plasma wave, validating Eq.~(\ref{eq:tbmt}) for PBAs.


Investigation of Eq.~(\ref{eq:tbmt}) in cylindrical coordinates provides a clear understanding of the physics of the spin precession for a single electron. In cylindrical coordinates, the spin vector is $\mathbf{s}=(s_r,s_{\phi},s_z)$, where $s_i=\mathbf{s}\cdot \mathbf{e_i}$, and where $\mathbf{e_i}$ correspond to the unit vectors in the radial ($\mathbf{e}_{r}$), azymuthal ($\mathbf{e}_{\phi}$), and longitudinal  ($\mathbf{e}_{z}$) directions. In addition, $(r,\phi,z)$ are the radial, azimuthal and longitudinal coordinates. To simplify the notation, each spin component is normalized to the absolute value of the electron spin, such that $\mathbf{s}$ takes values between -1 and 1. 

The electric field inside the plasma wave has both radial and longitudinal components $\mathbf{E}=E_r\mathbf{e}_r+E_z\mathbf{e}_z$, with $E_z\sim E_r$. Moreover, for electrons with $\gamma\gg 1$, $v_{r}\ll v_z \simeq 1$, which means that the first term on the right-hand-side of Eq.~(\ref{eq:tbmt}) coincides with the radial plasma focusing force $F_r$ felt by a relativistic electron:

\begin{equation}
\label{eq:bminuse}
\left(\mathbf{B}-\mathbf{v}\times\mathbf{E}\right) = \left(B_{\phi}-v_z E_r+v_r E_z \right)\mathbf{e}_{\phi}\simeq \left(B_{\phi}-E_r\right)\mathbf{e}_{\phi} \equiv F_r \mathbf{e}_{\phi},
\end{equation}
where $v_r$ and $v_z$ are the radial and longitudinal components of the electron velocity, and $-F_r$ is the radial force that acts on a relativistic electron in a plasma accelerator. Furthermore, the magnetic field of cylindrically symmetric plasma waves is purely azimuthal, i.e. $\mathbf{B}=B_{\phi} \mathbf{e}_{\phi}$~\cite{bib:lu_prl_2006}. Since for ultra-relativistic electrons $\gamma\gg 1$ (the most relevant scenario for PBAs), $v_{\phi}\ll v_z$, where $v_{\phi}$ is the azimuthal component of the electron velocity, then $\mathbf{v}\cdot \mathbf{B}=v_{\phi}B_{\phi}\ll 1$. Consequently, the last term on the right-hand-side of Eq.~(\ref{eq:tbmt}) can be neglected, and hence $F_r$ fully determines the spin precession of a single electron in PBAs.

Inserting Eq.~(\ref{eq:bminuse}) in Eq.~(\ref{eq:tbmt}), and neglecting the $\mathbf{v}\cdot\mathbf{B}$ term in Eq.~(\ref{eq:tbmt}) yields:

\begin{subequations}
\label{eq:spin}
\begin{align}
\frac{d s_{r}}{ d t} &= s_z \left(a+\frac{1}{\gamma}\right) F_r + s_{\phi} \dot{\phi} \label{eq:spina} \\ 
\frac{d s_{z}}{ d t}  &= -\left(a+\frac{1}{\gamma}\right) F_r s_r \label{eq:spinb}\\
\frac{d s_{\phi}}{ d t} &= -s_r \dot{\phi}, \label{eq:spinc}
\end{align}
\end{subequations}
where $\dot{\phi}\equiv \mathrm{d} \phi / \mathrm{d} t$.

Eqs.~(\ref{eq:spin}) represent a set of coupled differential equations which describe the spin precession of a single electron in the field structure of a PBA, valid as long as plasma waves are cylindrically symmetric. According to Eq.~(\ref{eq:spin}), the spin precession is mostly due to the anomalous electron magnetic moment when $\gamma\gg 1/a \simeq 10^3$. For $\gamma\ll 1/a$, the anomalous electron magnetic moment can be neglected. 

Solutions for Eqs.~(\ref{eq:spin}) can be retrieved by calculating the transverse electron trajectory in linear focusing and accelerating forces. Linear focusing forces are ideal for particle acceleration, as they preserve the transverse emittance of the accelerated beam. This is very important for HEP experiments which require low emittance beams in order to achieve high luminosities. Both the blowout, weakly- relativistic and even linear-regimes are characterized by linear focusing forces as long as the electron beam propagates close to the axis. Thus, $F_r=\mathbf{F}_{\perp}\cdot \hat{e}_{r}=\alpha \mathbf{x}_{\perp}\cdot \mathbf{x}_{\perp}/|\mathbf{x}_{\perp}|$, where $\mathbf{F}_{\perp}=(F_x,F_y)$ and $\mathbf{x}_{\perp}=(x,y)$. The parameter $\alpha=1/2$ if the blowout is complete. Otherwise, $\alpha<1/2$.

In the presence of uniform accelerating fields, $E_{\mathrm{accel}}$, the beam relativistic factor varies with $\gamma=\gamma_0+E_{\mathrm{accel}} t$, where $\gamma_0=\gamma(t=0)$. This simple model for the acceleration of electrons in plasma waves is accurate for a significant fraction of the acceleration distance, and only fails when $E_{\mathrm{accel}}$ is significantly reduced due to the laser pulse pump depletion in LWFAs, or electron beam head-erosion in PWFAs, and when the acceleration distance is comparable with the dephasing length in both LWFAs and PWFAs.

Under the conditions mentioned above, the radial electron oscillations in the plasma wave ion channel are given by~\cite{bib:glinec_epl_2008}:

\begin{equation}
\label{eq:glinec}
(\gamma_0+E_{\mathrm{accel}} t) \ddot{\mathbf{x}}_{\perp}(t)+\gamma_0 \dot{\mathbf{x}}_{\perp}(t) + \alpha \mathbf{x}_{\perp}(t) = 0,
\end{equation}
where $\mathbf{x}_{\perp0}=(x_0,y_0)$ is the initial transverse position of the electron. Eq.~(\ref{eq:glinec}) can be solved analytically for an electron initially located at $\mathbf{x}_{\perp0}$, with initial transverse momentum $\mathbf{p}_{\perp0}=\gamma \mathbf{v}_{\perp0}$, for which $\sqrt{\alpha \gamma_0}/E_{\mathrm{accel}}\gg 1$ and $\sqrt{\alpha \gamma}/E_{\mathrm{accel}}\gg 1$, yielding: 

\begin{eqnarray}
\label{eq:perp_traj}
\mathbf{x}_{\perp}(t) & = & \mathbf{x}_{\perp0} \left(\frac{\gamma_0}{\gamma(t)}\right)^{1/4} \cos\left(\frac{2\sqrt{\alpha \gamma(t)}}{E_{\mathrm{accel}}} - \frac{2\sqrt{\alpha \gamma_0}}{E_{\mathrm{accel}}}\right)+\nonumber \\
& + & \frac{\mathbf{p}_{\perp0}}{\left(\alpha^2\gamma\gamma_0\right)^{1/4}} \sin\left(\frac{2\sqrt{\alpha \gamma(t)}}{E_{\mathrm{accel}}} - \frac{2\sqrt{\alpha \gamma_0}}{E_{\mathrm{accel}}}\right).
\end{eqnarray}

Provided that $\mathbf{x}_{\perp0}$ and $\mathbf{p}_{\perp0}$ are known, Eqs.~(\ref{eq:spin}) and (\ref{eq:perp_traj}) fully describe the spin precession dynamics of a single electron in a PBA. 

The most interesting limit for HEP applications is associated with low beam emittances, characterized by low average $|\mathbf{p}_{\perp0}|$ in comparison to the average $|\mathbf{x}_{\perp0}|$, such that the eccentricity of the electrons trajectory in the plane perpendicular to the propagation direction is much higher than one. According to Eq.~(\ref{eq:perp_traj}) this occurs for $\langle|\mathbf{x}_{\perp0}|\rangle\gg \langle |\mathbf{p}_{\perp0}|\rangle /\sqrt{\alpha \gamma_0}$, where $\langle|\mathbf{x}_{\perp0}|\rangle\simeq \sigma_r$ is the typical transverse width of the electron beam, and where $\langle |\mathbf{p}_{\perp0}|\rangle$ is the typical electron beam perpendicular momentum. Equivalently, the effects due to finite beam emittance can be neglected if the normalized beam emittance, $\epsilon_{N}$, is much lower than
\begin{equation}
\label{eq:emittance}
\epsilon_{N}^{\mathrm{max}}[\mathrm{mm} \cdot \mathrm{mrad}] \ll 188 \sqrt{\alpha} \left(\frac{\hat{\sigma}_r}{10~\mu\mathrm{m}}\right)^2\left(\frac{E_0}{10~\mathrm{GeV}}\right)^{1/2}\left(\frac{n_0}{10^{16}~\mathrm{cm}^{-3}}\right)^{1/2},
\end{equation}
where the normalized emittance is estimated according to $\epsilon_{N}=\langle|\mathbf{x}_{\perp}|\rangle \langle|\mathbf{p}_{\perp}|\rangle$, and where $E_0$ is the initial beam energy [we note that $\epsilon_{\mathrm{N}}=\epsilon_{\mathrm{N}}^{\mathrm{max}}$ corresponds to matched PWFA propagation regimes which minimize the betatron oscillations of the driving electron bunch]. These analytical results can still be applied to unmatched regimes in the PWFA or in configurations associated with the external injection of electrons in plasma wakes. In fact, the beam energy, energy spread, emittance, transverse spot size, and number of accelerated electrons can only be effectively controlled when electrons are externally injected at the rear of the plasma wave~\cite{bib:lu_prstab_2007,bib:huang_pac_2009}. In this case it is advantageous to reduce the beam emittance to be as small as possible, thereby increasing the luminosity of the bunches, entering in a regime where Eq.~(\ref{eq:emittance}) is valid. Later in this work we will show that this condition is also advantageous for beam polarization conditioning in plasma accelerators, and that the depolarization of lower emittance beams is lower in comparison to higher emittance bunches.

In the conditions of Eq.~(\ref{eq:emittance}), the electron trajectories are almost planar, and $d s_{\phi}/dt\ll 1$ which implies that $s_{\phi}$ and $s_r^2+s_z^2=1-s_{\phi 0}^2$ are conserved, with $s_{\phi0}$ the initial tangential spin component. Under these conditions, the expression for the longitudinal component of the spin is found by combining Eqs.~(\ref{eq:spina}) and (\ref{eq:spinb}), yielding:

\begin{equation}
\label{eq:slong2}
s_z(t) = \sqrt{1-s_{\phi 0}^2} \sin\left[-\int_{0}^{t} \left(a+\frac{1}{\gamma}\right)F_r dt + \arctan\left(\frac{s_{z0}}{s_{\phi0}}\right)\right], 
\end{equation}
where $s_{z0}$, and $s_{r0}$ are the initial longitudinal and radial spin components, and where the radial component of the spin is given by $s_r^2(t)=1-s_{\phi0}^2-s_z^2(t)$. Eq.~(\ref{eq:slong2}), which is valid for arbitrary $F_r$ as long as Eq.~(\ref{eq:emittance}) is satisfied, describes the spin precession of a single electron in the PBA. 

\begin{figure}[thbp]
\begin{center}
\includegraphics[scale=0.75]{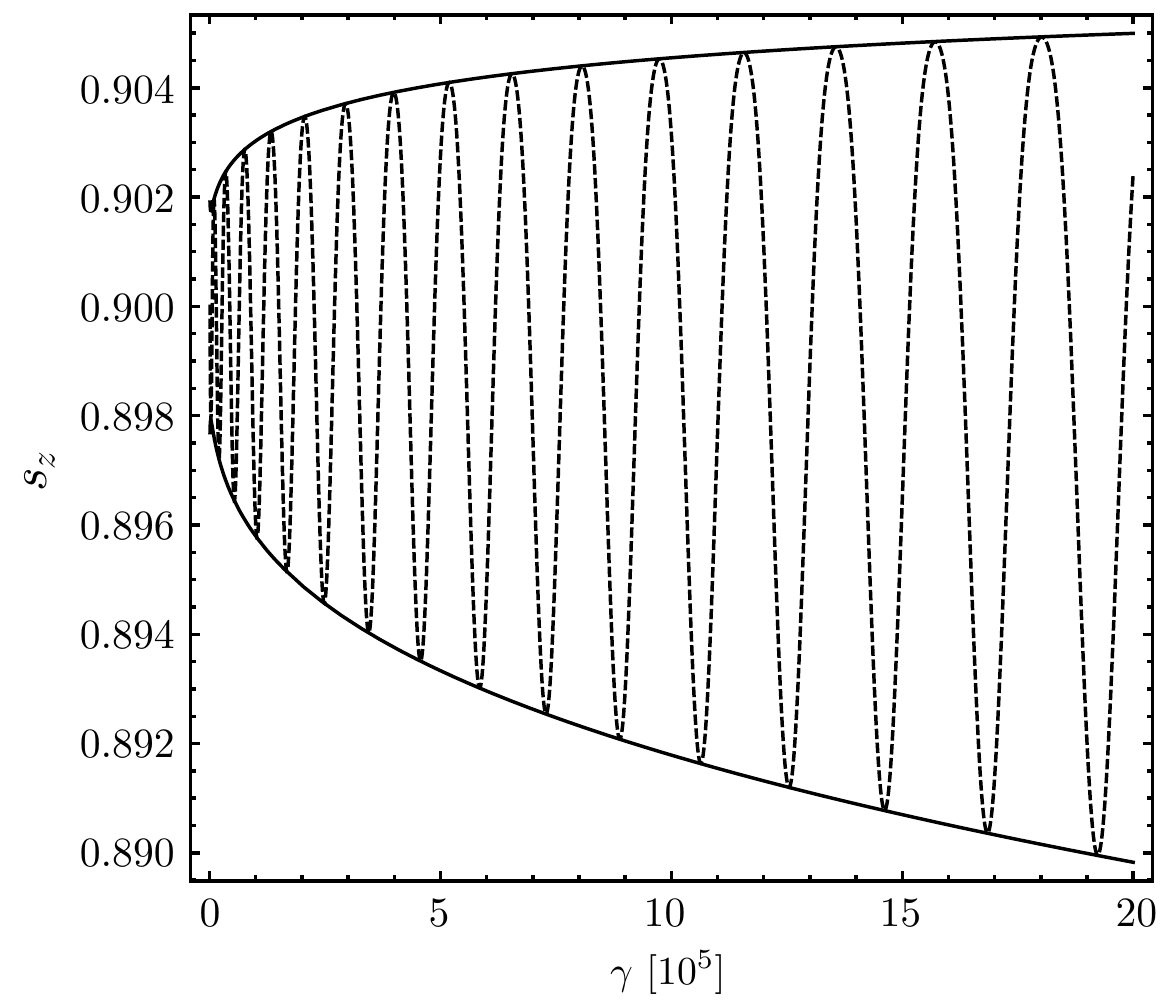}
\caption{\label{fig:szmodel} Evolution of the longitudinal spin component $s_z$ as a function of $\gamma$ for a single electron with zero initial velocity, with $r_0=0.5~c/\omega_p$, $\gamma_0=10^3$, $s_{z0}=0.9$, and $s_{r0}=0.1$. The solid lines denote the envelope of the oscillations. The dashed lines represent $s_z(t)$. The amplitude of the oscillation is on the order of $s_{z0}/100$ for an electron accelerating to 1 TeV, indicating the potential of plasma based accelerators to accelerate polarized electron beams. The period of the oscillation of $s_z$ is equal to the betatron oscillations period.}   
\end{center}
\end{figure}

The radial plasma focusing force in Eq.~(\ref{eq:slong2}) depends on the electron trajectory as $F_r=F_r(r(t))$. The radial electron trajectory for low emittance electron beams ($\langle |\mathbf{p}_{\perp0}| \rangle \ll \sqrt{\alpha\gamma} \langle |\mathbf{x}_{\perp0} | \rangle$) is obtained from Eq.~(\ref{eq:perp_traj}) by replacing $\mathbf{x}_{\perp}(t)$ by $r(t)$, and $\mathbf{x}_{\perp0}$ by $(x_0 |x_0| + y_0 |y_0|)/\sqrt{x_0^2+y_0^2}\equiv r_0$. Thus Eq.~(\ref{eq:slong2}) can be re-written as:

\begin{equation}
\label{eq:slong3}
s_z(\gamma(t)) =\sqrt{1-s_{\phi 0}^2} \sin\left[r_0 \Phi(\gamma(t)) + \arctan\left(\frac{s_{z0}}{s_{r0}}\right)\right],
\end{equation}
where $\Phi(\gamma(t))=-[(1+a\gamma) (\alpha^2 \gamma_0/\gamma^3)^{1/4}] \sin\left[2\sqrt{\alpha}(\sqrt{\gamma}-\sqrt{\gamma_0})/E_{\mathrm{accel}}\right]$. Equation~(\ref{eq:slong3}) gives the evolution of the longitudinal spin precession component for a single electron that propagates under linear focusing forces and constant accelerating gradients. It shows that $s_z$ oscillates with the betatron frequency, and that the amplitude of the oscillations depends directly on the electron energy, since $\Phi$ is a function of $\gamma$. Fig.~\ref{fig:szmodel} illustrates the evolution of $s_z$ for a single electron according to Eq.~(\ref{eq:slong3}) with $r_0=0.5~c/\omega_p$, $\gamma_0=10^3$, $s_{z0}=0.9$, and $s_{r0}=0.1$, showing that the spin precession amplitude is on the order of $s_{z0}/100$ thus suggesting that plasma accelerators can be efficiently used to accelerate polarized electron beams in conditions that are relevant for high energy physics experiments.

\begin{figure}[thbp]
\begin{center}
\includegraphics[scale=0.4]{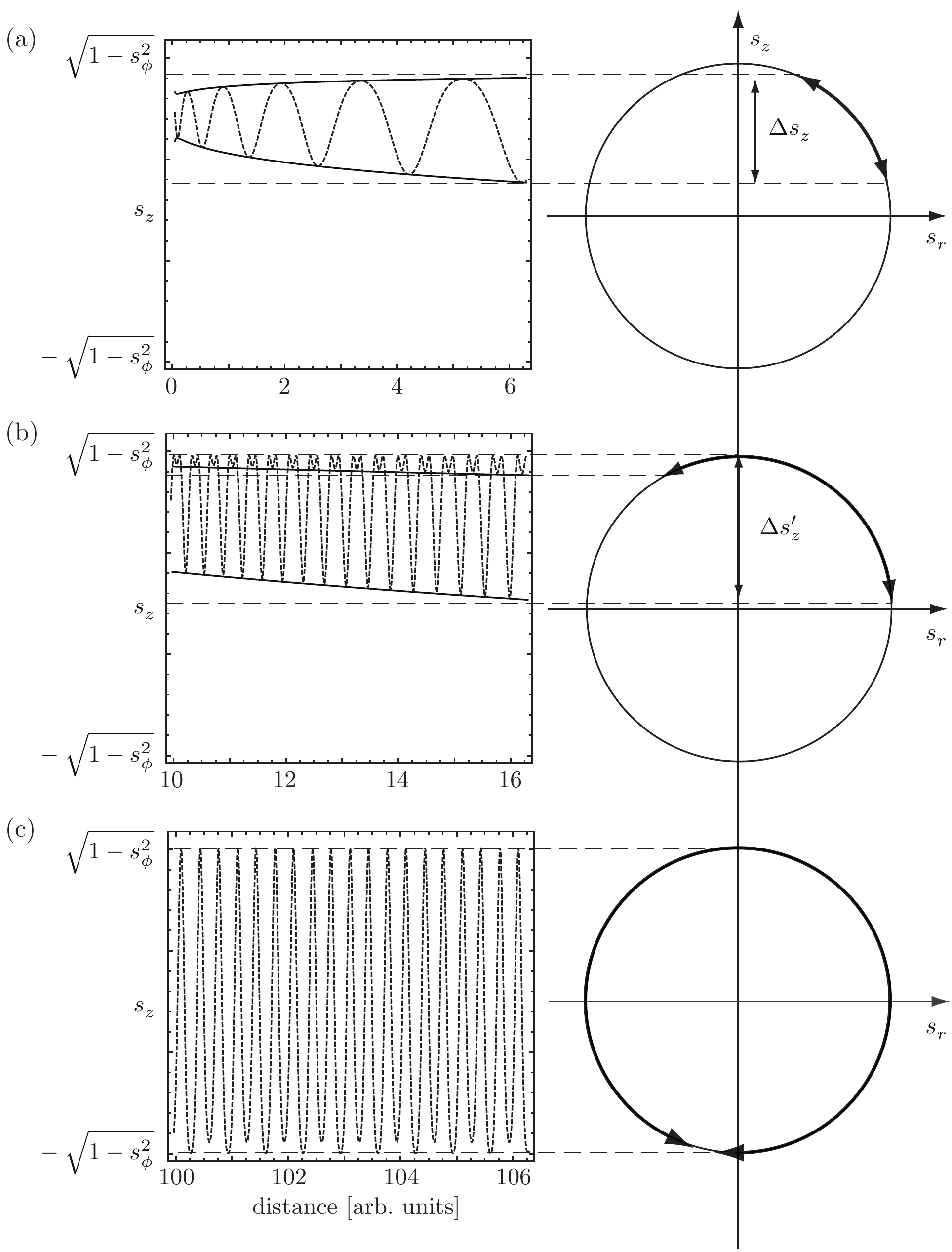}
\caption{\label{fig:model} Qualitative temporal evolution of $s_z$ (solid-gray lines) during acceleration. (a) shows $s_z(t)$ for for $\gamma<\gamma_1$, (b) for $\gamma_1<\gamma<\gamma_2$, and (c) for $\gamma>\gamma_2$. The solid-black lines are $s_z^{\mathrm{max/min}}$. The circumferences show the possible trajectories of the spin in the $(s_z,s_r)$ phase space for constant $s_{\phi}$. The arches on top represent the trajectory $(s_z,s_r)$ for the time frame correspondent to the left plots.}   
\end{center}
\end{figure}

The qualitative evolution of $s(z)$ given by Eq.~(\ref{eq:slong3}) is represented in Fig.~\ref{fig:model}. If the spin of the electron initially lies in the first or second quadrants of the $(s_r,s_z)$ phase space (i.e. if $-\pi<\arctan\left(s_{z0}/s_{r0}\right)<0$), $s_z$ reaches its maximum global extremum, $s_z = \sqrt{1-s_{\phi0}^2}$, when $\gamma$ is higher than

\begin{equation}
\label{eq:slong3max}
\gamma >\gamma_1 = \frac{\left[2 \arctan{\left(s_{z0}/s_{r0}\right)}-\pi\right]^4}{16 a^4 r_0^4 \alpha^2 \gamma_0},
\end{equation}
and it reaches its minimum global extremum, $s_z = -\sqrt{1-s_{\phi0}^2}$, when $\gamma$ is higher than
\begin{equation}
\label{eq:slong3min}
\gamma > \gamma_2 = \frac{\left[2 \arctan{\left(s_{z0}/s_{r0}\right)}+\pi\right]^4}{16 a^4 r_0^4 \alpha^2 \gamma_0},
\end{equation}
If the spin of the electron initially lies in the third or fourth quadrants of the $(s_r,s_z)$ phase-space, i.e. if $\pi>\arctan\left(s_{z0}/s_{r0}\right)>2 \pi$, then $\gamma_1$ ($\gamma_2$) is replaced by $\gamma_2$ ($\gamma_1$). Inserting typical values for $r_0\sim c/\omega_p$ and for $\gamma_0\sim 10^3$ in Eqs.~(\ref{eq:slong3max}) and (\ref{eq:slong3min}) yields $\gamma_1 \sim 10^{9}$ and $\gamma_2 \sim 10^{11}$. For 1~TeV electrons where $\gamma\sim 10^{6}$, this shows that the spin vector never performs one complete revolution in the $(s_r,s_z)$ phase space during the acceleration in PBAs. This feature is in contrast with circular accelerators, where the spin revolution frequency is of the same order of the orbital frequency of the electron beam~\cite{bib:mane_rpp_2005}. 

The low amplitudes of oscillation for $s_z(t)$ and $s_r(t)$, are due to the low values that $\Phi(t)$ takes for typical PBA parameters. For $\Phi(t)\ll 1$, Eq.~(\ref{eq:slong3}) becomes:

\begin{equation}
\label{eq:s3_approx}
s_z(t) =s_{z0}+ s_{r0} r_0 \Phi(t)-\frac{s_{z0} r_0^2 \Phi(t)^2}{2}+\mathcal{O}[\Phi(t)^3],
\end{equation}
which reveals that $s_z$ oscillates between the maximum $s_z^{\mathrm{max}}$ and the minimum $s_z^{\mathrm{min}}$ given by: 
\begin{equation}
\label{eq:s3_approx_max}
s_z^{\mathrm{max/min}}(t) = s_{z0} \pm s_{r0} r_0 \sqrt{\alpha} \left(\gamma \gamma_0\right)^{1/4}-\frac{s_{z0} r_0^2 \alpha \left(\gamma \gamma_0\right)^{1/2}}{2}, 
\end{equation}
with a corresponding oscillation amplitude of:
\begin{equation}
\label{eq:amplitude_sz}
\Delta s_z \simeq 2 r_0 s_{r0} \sqrt{\alpha} \left(\gamma \gamma_0\right)^{1/4} a.
\end{equation}
Note that Eq.~(\ref{eq:amplitude_sz}) is valid as long as $s_{z0}$ or $s_{r0}$ remain in the same quadrant, i.e., $\gamma<\gamma_1$. For $\gamma>\gamma_1$, it is modified to $\Delta s_z^{\prime}=\Delta s_z+\delta$, where $\delta=\pm \sqrt{1-s_{\phi0}^2} \mp s_z^{\mathrm{max/min}}$ (Fig.~\ref{fig:model}).

The evolution of $\Delta s_z$ generally depends on the initial amplitudes of the betatron oscillations (through $r_0$), on the reduction of the betatron oscillations amplitude as the electron accelerates (through $\gamma$), on the initial electron energy (through $\gamma_0$), and on the focusing force (through $\alpha$). The spin precession amplitude $\Delta s_z$ is lower for lower $r_0$, i.e., for beam electrons that are closer to the axis. In addition, $\Delta s_z$ increases throughout the acceleration, since the integral in Eq.~(\ref{eq:slong2}) increases as the betatron amplitude of oscillation of single beam electrons is reduced during the acceleration~\cite{bib:glinec_epl_2008}. If the betatron oscillations amplitude would be constant throughout the acceleration, the average value of $\Delta s_z$ over a multiple of the betatron period would be constant since $\langle\Delta s_z\rangle\propto\int_0^t (a+\gamma^{-1}) F_r dt=0$. Therefore, at the later stages of the acceleration, when the acceleration gradients are more significantly reduced, and when the variations of the amplitude of the betatron oscillations are thus more noticeable, $d\langle\Delta s_z\rangle/d t$ also decreases. Lower $\alpha$'s that correspond to weaker focusing forces also reduce $\Delta s_z$. Therefore, the linear regime of plasma based acceleration can improve the polarization conditioning in plasma accelerators in comparison to the blowout regime. This requires, however, very narrow beams, such that the electron beam responds only to linear focusing forces.


\section{Beam polarization}
\label{sec:polarization}

The beam depolarization rates are critical to assess the potential of PBA for HEP experiments. The electron beam polarization component in a given spatial direction is given by the statistical average of $\mathbf{s}$ weighted by the polarized electron beam distribution function $f_e$~\cite{bib:mane_rpp_2005}: 

\begin{equation}
\label{eq:polarization}
\mathbf{P} = \int f_e \mathbf{s} \mathrm{d}\mathbf{x}\mathrm{d}\mathbf{v}\mathrm{d}\mathbf{s}\Big{/}\int f_e \mathrm{d}\mathbf{x}\mathrm{d}\mathbf{v}\mathrm{d}\mathbf{s},
\end{equation}
We assume that the distribution function of the accelerated electron beam can be separated by its spatial ($\mathrm{R}(\mathbf{r})$), velocity ($\mathbf{V}(\mathbf{v})$), and spin ($\mathbf{S}(\mathbf{s})$) distributions as:

\begin{equation}
\label{eq:beam_general}
f_{e}=\mathrm{R}(\mathbf{r}) \mathrm{V}(\mathbf{v}) \mathrm{S} (\mathbf{s}).
\end{equation}
A transversely cylindrically symmetric Gaussian density distribution is considered, $\mathrm{R}(\mathbf{r})=n_{b0} \exp[-2 r^2/\sigma_r^2] \mathrm{Z}(z)$, with $n_{b0}$ being the peak electron beam density, $\sigma_r$ the electron beam transverse width, and where $\mathrm{Z}(z)$ is chosen to provide ideal beam loading~\cite{bib:tzoufras_prl_2008} which guarantees that $E_z$ can be the same throughout the beam. In fact, by choosing $\mathrm{Z}(z)$ according to~\cite{bib:tzoufras_prl_2008}, all beam electrons accelerate with the same accelerating gradient, and thus $\mathbf{P}$ is independent of $z$. For a zero-emittance, and zero-energy spread electron beam, $\mathbf{V}(\mathbf{v})$ is given by $\mathbf{V}(\mathbf{v})=\delta(\mathbf{v}_{\perp}) \delta(v_z-v_{z0})$, where $v_{z0}=(\gamma_0^2-1)^{1/2}/\gamma_0$ is the initial beam longitudinal velocity. The initial electron beam distribution function becomes fully characterized by choosing $\mathrm{S}(\mathbf{s})$; for the sake of simplicity, an uniformly polarized electron beam with $\mathrm{S}(\mathbf{s})=\delta(\mathbf{s}-\mathbf{S_0})$ is assumed. Beams with arbitrary polarization can be examined by allowing that $\mathbf{S}_0\leq1$, where $|\mathbf{S}_0|$ coincides with the initial polarization of the beam $\mathbf{P}_0$. Physically this means that the average spin of the beam electrons located between $\mathbf{r}$ and $\mathbf{r}+\delta\mathbf{r}$, and between $\mathbf{v}$ and $\mathbf{v}+\delta\mathbf{v}$ can be less than unity. Thus, the electron beam distribution function becomes:

\begin{equation}
\label{eq:beam}
f_{e}=n_{b}\exp{\left[-\frac{2 r_0^2}{\sigma_r^2}\right]} \mathrm{Z}(z) \delta(\mathbf{v}_{\perp})\delta(v_z-v_{z0}) \delta(\mathbf{s}-\mathbf{S_0}).
\end{equation}

The distribution function considered in Eq.~(\ref{eq:beam}) which can be used as long as the beam emittance and energy spread is negligible, illustrates the key properties of spin precession in PBAs.


The calculation of the beam polarization requires cartesian coordinates because the cylindrical coordinate system is attached to the trajectory of each beam electron. Thus, in order to retrieve the beam polarization, $(s_r,s_{\phi},s_z)$ must be projected in the cartesian coordinates basis, according to:

\begin{equation}
\label{eq:sx}
s_x= s_r \cos({\phi})-s_{\phi0} \sin(\phi) = \pm \sqrt{1-s_z^2-s_{\phi0}^2} \cos(\phi)-s_{\phi0} \sin{(\phi)}
\end{equation} 

\begin{equation}
\label{eq:sy}
s_y= s_{\phi0} \cos({\phi})+s_{r} \sin{(\phi)} = s_{\phi0} \cos({\phi})\pm \sqrt{1-s_z^2-s_{\phi0}^2} \sin{(\phi)}, 
\end{equation} 
being $s_z$ identical in both coordinate systems. 

Analytical results for the beam polarization can be retrieved in the limit $\Phi\ll 1$, combining Eq.~(\ref{eq:slong3}), Eq.~(\ref{eq:polarization}), and Eqs.(\ref{eq:beam}-\ref{eq:sy}), yielding:

\begin{equation}
\label{eq:p_perp}
\frac{\mathbf{P}_{\perp}}{P_0} = \mathbf{S}_{\mathrm{0{\perp}}} \left[1 - \frac{\Phi^2 \sigma_r^2}{8} - \frac{\Phi^4 \sigma_r^4 }{96 }+\mathcal{O}(\Phi^5)\right],
\end{equation}

\begin{equation}
\label{eq:p_3}
\frac{P_z}{P_0} = S_{\mathrm{0z}} \left[1 - \frac{\Phi^2 \sigma_r^2}{4} + \frac{\Phi^4 \sigma_r^4}{48} +\mathcal{O}(\Phi^5)\right],
\end{equation}
where $\mathbf{P}_{\perp}=(P_x,P_y)$ is the polarization in each transverse direction, $P_z$ the polarization along the $z$ direction, $\mathbf{S}_{\mathrm{0{\perp}}}=\mathbf{P}_{\perp}(t=0)$, and $S_{0z}=P_{z}(t=0)$. Comparison between Eqs.~(\ref{eq:p_perp}) and Eq.~(\ref{eq:p_3}) reveals that $|\dot{\mathbf{P}}_{\perp}|\simeq (1/2) |\dot{P}_z|$. This indicates that the conditioning of transversely polarized electron beams with zero emittance is more effective than the polarization conditioning of longitudinally polarized electron beams. The total polarization of the beam $P=\sqrt{\mathbf{P}_{\perp}^2+P_z^2}$ is given by: 

\begin{equation}
\label{eq:p_total}
\frac{P}{P_0}=1-\frac{\sigma_r^2 (1+s_{z0}^2) \Phi^2}{8}+\frac{\sigma_r^4 (4+7 s_{z0}^2-3 s_{z0}^4) \Phi^4}{384}+\mathcal{O}(\Phi^5),
\end{equation}
and confirms the aforementioned conclusion that the conditioning of transversely polarized zero-emittance electron beams is more effective since $\dot P/P_0$ is minimum for $s_{z0}=0$. Eq.~(\ref{eq:p_perp}) also reveals that the direction of $\mathbf{P}_{\perp}$ is preserved during the propagation, since $P_{x}/P_{y}=s_{x0}/s_{y0}$ remains constant. The polarization vector then oscillates in the plane defined by $\mathbf{s}_{\perp0}$ and $\mathbf{e}_z$. Similarly to the single electron spin precession dynamics, $\mathbf{P}$ never completes a revolution in the $(\mathbf{P}_{\perp},P_{z})$ plane, and, in fact, the depolarization amplitude is lower than the spin precession amplitude, since $|\dot{\mathbf{P}}|\simeq \dot{\mathbf{s}}^2 \ll 1$. This reveals that the beam polarization can be maximized in the same manner that the spin precession amplitudes are minimized (cf. Sec.~\ref{sec:precession}).

\begin{figure}[htbp]
\begin{center}
\includegraphics[scale=0.7]{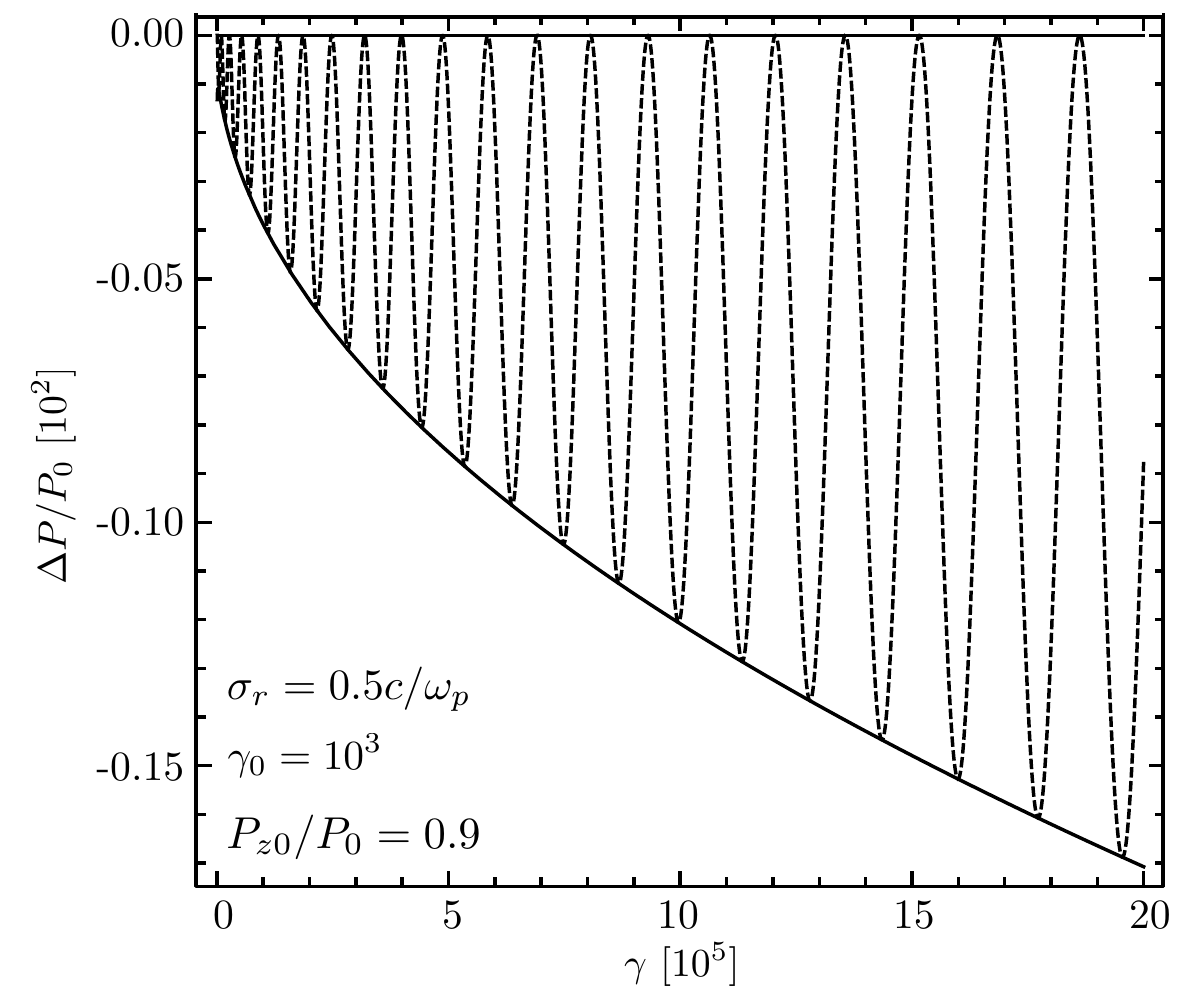}
\caption{\label{fig:model2} Temporal evolution of the total polarization of a beam which accelerates from 0.5 GeV ($\gamma_0=10^3$) to 1 TeV ($\gamma_0=2\times 10^6$), considering $P_{z0}/P_0=0.9$. The solid black lines are the upper and lower envelopes $\Delta P/P_0$. The dashed lines represent the evolution of $\Delta P/P_0$.}   
\end{center}
\end{figure}

Fig.~\ref{fig:model2} illustrates the qualitative evolution of $\mathbf{P}$, showing that the polarization oscillates with twice the betatron frequency between $P_0$ and minimum values with an amplitude given by:

\begin{equation}
\label{eq:p_perp_amp}
\frac{|\Delta \mathbf{P}_{\perp}|}{P_0}=\frac{s_{\perp0} \sigma_r^2 \left(\alpha \gamma_0 \gamma\right)^{1/2} a^2}{8},
\end{equation}

\begin{equation}
\label{eq:p_3_amp}
\frac{|\Delta P_{z}|}{P_0}=\frac{s_{z0}\sigma_r^2 \left(\alpha \gamma_0 \gamma\right)^{1/2} a^2}{4},
\end{equation}

\begin{equation}
\label{eq:ptotal_amp}
\frac{|\Delta {P}|}{P_0}=\frac{\left(1+s_{z0}^2\right) \sigma_r^2 \left(\alpha \gamma_0 \gamma\right)^{1/2} a^2}{8},
\end{equation}
where $\Delta\mathbf{P}_{\perp}=(\Delta P_x,\Delta P_y)$ is the depolarization amplitude in the transverse directions, and $\Delta P$ is the total beam depolarization amplitude. Eqs.~(\ref{eq:p_perp_amp}), (\ref{eq:p_3_amp}), and (\ref{eq:ptotal_amp}) predict the maximum depolarization for an accelerated electron beam with a final energy $E=m_e c^2 \gamma$ at the end of the plasma accelerator. It is important to note that Eqs.~(\ref{eq:p_perp_amp}), (\ref{eq:p_3_amp}), and ($\ref{eq:ptotal_amp}$) imply that PBAs can be designed to keep the beam polarization exactly equal to the initial polarization because for higher energy beams, with longer betatron wavelengths, it is conceivable that the length of the plasma can be adjusted to a multiple of half the betatron wavelength guaranteeing that $\mathbf{P} \simeq \mathbf{P}_0$.

\section{Self-guided vs. Externally-guided LWFAs}
\label{sec:sgvseg}

Two regimes for stable propagation have been proposed in the blowout regime of LWFA, either using external guiding structures (such as parabolic plasma channels), or by adjusting the laser and plasma parameters to ensure self-guiding~\cite{bib:lu_prstab_2007}. It is straightforward to retrieve the total depolarization associated with externally-guided or self-guided LWFAs of similar output energy according to the scaling laws from Ref.~\cite{bib:lu_prstab_2007}. It was assumed in this calculation that the physical dimensions of the beam are the same for the externally-guided or the self-guided scenarios, while the rest of the parameters such as laser energy, and plasma density, were chosen to ensure stable accelerating structures. Denoting the total depolarization associated with the self-guided propagation regime by $\Delta P^{\mathrm{sg}}/P_0$, and the depolarization associated with the externally-guided propagation regime by $\Delta P^{\mathrm{eg}}/P_0$, Eq.~(\ref{eq:ptotal_amp}) leads to:

\begin{equation}
\label{eq:self_guided}
\frac{\Delta P^{\mathrm{sg}}}{P_0} \simeq 1.17\times 10^{-4} \frac{\left(1+s_{z0}^2\right) \sqrt{\gamma_0}}{\gamma^{1/3}} \left(\frac{\hat{\sigma}_r}{10~\mu\mathrm{m}}\right)^2 \left(\frac{1\mu\mathrm{m}}{\hat{\lambda}_0}\right)^2,
\end{equation}

for the self-guided propagation regime, and as

\begin{equation}
\label{eq:external_guided}
\frac{\Delta P^{\mathrm{eg}}}{P_0} \simeq 8.9\times10^{-5}\frac{\left(1+s_{z0}^2\right) \sqrt{\alpha \gamma_0}}{ \gamma^{1/2}} \left(\frac{\hat{\sigma}_r}{10~\mu\mathrm{m}}\right)^2 \left(\frac{1\mu\mathrm{m}}{\hat{\lambda}_0}\right)^2,
\end{equation}
for the externally-guided propagation regime, where $\hat{\lambda}_0$ is the laser central wavelength in $\mu$m, and $\hat{\sigma}_r$ is the electron beam width measured in $\mu$m. Although the depolarization increases during the acceleration, Eqs.~(\ref{eq:self_guided}) and (\ref{eq:external_guided}) reveal that higher energy matched propagation LWFAs lead to lower depolarizations. This is because a higher energy matched LWFA uses lower densities, lowering the normalized $\sigma_r$ (the beam dimensions in physical units are kept constant), and thus $\Delta P/P_0$. The focusing force is also lower for externally guided propagation regimes, which further decreases the depolarization rates. In standard accelerators, however, $\Delta P/P_0$ increases for higher energies. This result thus emphasizes an important advantage to accelerate highly polarized electron beams to very high energies in PBA.

The results of Eqs.~(\ref{eq:self_guided}) and (\ref{eq:external_guided}) can be illustrated by considering typical electron beam parameters. At SLAC, $\hat{\sigma}_r=10~\mu\textrm{m}$, and $\gamma_0=6\times10^4$. Thus, the maximum total depolarization associated with a $0.5$ TeV accelerator is $\Delta P^{\mathrm{sg}}/P_0=0.16 \%$ and $\Delta P^{\mathrm{eg}}/P_0=0.025 \%$, where we used $\alpha=0.35$, estimated from 3D PIC simulations~\cite{bib:vieira_ieee_2008}. We note that these results do not take into account the beam depolarization due to the beam-beam interaction at the interaction point. Nevertheless, it is instructive to compare the typical depolarizations expected in PBAs with the expected depolarization at the ILC, which is on the order of $\Delta P^{\mathrm{ILC}}\simeq 0.1-0.2$~\cite{bib:pick_epac_2006} [we note that $\Delta P^{\mathrm{ILC}}$ is mostly due to the depolarization at the beam-beam collision point]. Although this work only takes into account the beam depolarization during the acceleration, this comparison thus reveals that plasma accelerators should also be able to fulfill the depolarization requirements of HEP experiments.

\section{Numerical Simulations}
\label{sec:simulations}

In order to investigate the acceleration of polarized electron beams in realistic conditions, the T-BMT equations are further explored with a numerical spin-tracking algorithm. The evolution of the polarization is firstly examined using prescribed fields and electron trajectories (cf. Sec.~\ref{sec:precession}), where very fast parameter scans can be performed even for $> 1~\mathrm{TeV}$ accelerators. Then, the results from one-to-one 3D QuickPIC~\cite{bib:huang_jcp_2006} simulations are presented, providing strong evidences for PBA with low depolarization, in agreement with the theoretical predictions.

The similarities between Eq.~(\ref{eq:tbmt}) and the magnetic field component of the Lorentz force suggest that a non-relativistic Boris-pusher~\cite{bib:birdsall} can be used to solve the T-BMT equations. The rotation of the spin is determined by $\mathbf{\Omega}$, in the same way the rotation of the electron velocity is determined by $\mathbf{B}$ in the Lorentz force equation. For $\Delta t/\lambda_{\beta}\ll 1$, the spin rotates by an angle $\delta$, given implicitly by $\mathbf{d} = \mathbf{\Omega}\tan\left(\delta/2\right) \simeq \mathbf{\Omega} \Delta t/2$. The rotation of $\mathbf{s}$ between each time step is given by:

\begin{equation}
\label{eq:1ststep}
\mathbf{s}^{\prime}_{i} = \mathbf{s}_i + \mathbf{s}_i \times \mathbf{d},  
\end{equation}

and

\begin{equation}
\label{eq:2ndstep}
\mathbf{s}_{i+1} = \mathbf{s}_i + \frac{2 \mathbf{s}_i^{\prime} \times \mathbf{d}}{1+|\mathbf{d}|^2},  
\end{equation}
where $i$ is the iteration number. Eqs.~(\ref{eq:1ststep}) and (\ref{eq:2ndstep}) guarantee that $|\mathbf{s}_i|^{2}=|\mathbf{s}_{i+1}|^{2}$, providing an accurate numerical resolution of the T-BMT equations, to second order in $\Delta t$. 

In order to compare the analytical model with Eqs.~(\ref{eq:1ststep}) and (\ref{eq:2ndstep}), we consider prescribed radial focusing forces, $\mathbf{F}_{\perp}=\alpha \mathbf{x}_{\perp}$, and the electron trajectories described by Eq.~(\ref{eq:perp_traj}). It then follows that

\begin{eqnarray*}
\Omega_x & = & -\left(a+\frac{1}{\gamma(t)}\right) \left(E_x-B_y\right)\\
& = & -\alpha \left(a+\frac{1}{\gamma(t)}\right)  x(t),
\end{eqnarray*} 
and 
\begin{eqnarray*}
\Omega_y & = & \left(a+\frac{1}{\gamma(t)}\right) \left(E_y-B_x\right) = \\
& = & \alpha  \left(a+\frac{1}{\gamma(t)}\right) y(t). 
\end{eqnarray*}

\begin{figure}[htbp]
\begin{center}
\includegraphics[width=\columnwidth]{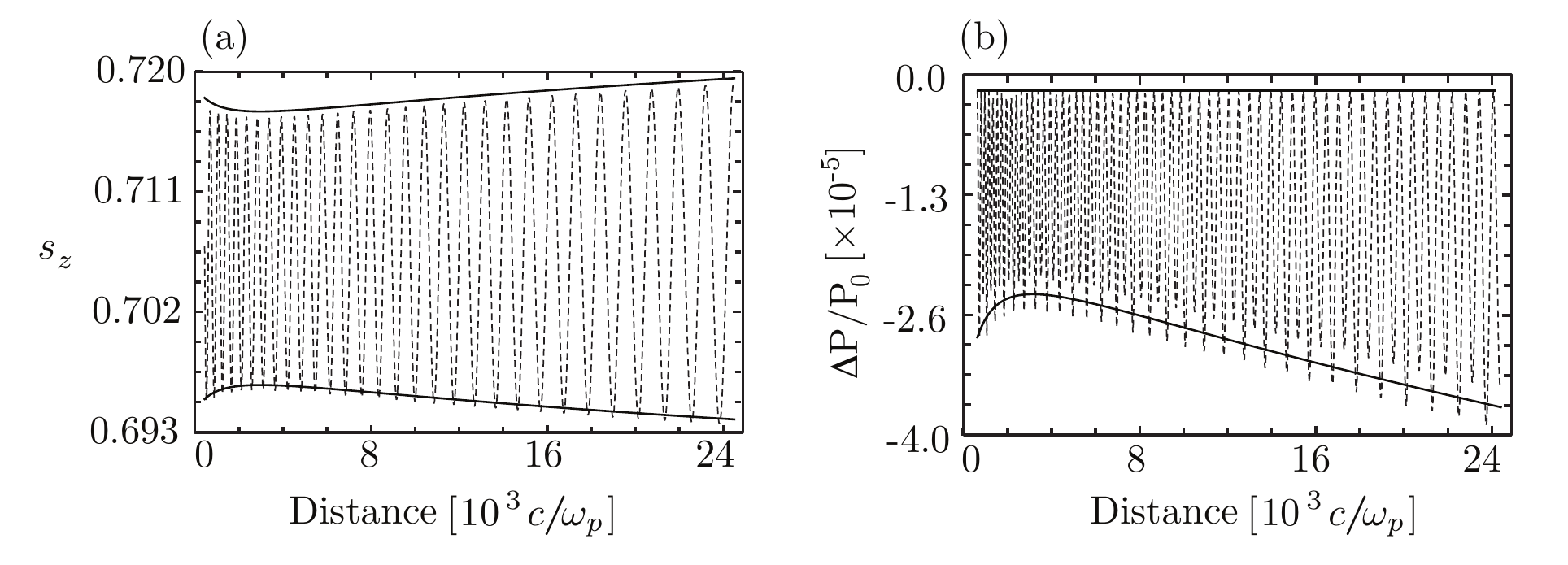}
\caption{\label{fig:spin_numerical} Comparison between the analytical model and the numerical resolution of the T-BMT equations using prescribed fields. (a) Longitudinal component of the spin evolution for a beam with $\sigma_r=0.1$, which accelerates in a constant accelerating gradient $E_{\mathrm{accel}}=0.6$, with $\alpha=0.35$. (b) Evolution of the total depolarization of the beam. The solid line corresponds to the analytical result, while the dashed line corresponds to the numerical solution.} 
\end{center}
\end{figure}

Fig.~\ref{fig:spin_numerical} shows a comparison between $s_z^{\mathrm{min/max}}$ (see Eq.~(\ref{eq:s3_approx_max})), $\Delta P/P_0$ (see Eq.~(\ref{eq:p_total})), and the numerical resolution of the T-BMT Equations (Eqs.~(\ref{eq:1ststep}) and (\ref{eq:2ndstep})). A zero-emittance electron beam was initialized with $\sigma_r=0.1~c/\omega_p$ and $\gamma_0=10^3$, which accelerates with $E_{\mathrm{accel}}=0.6$ during $L = 2.4 \times 10^{4}~c/\omega_p$. The numerical results are in excellent agreement with the analytical predictions.

\begin{figure}[htbp]
\begin{center}
\includegraphics[width=\columnwidth]{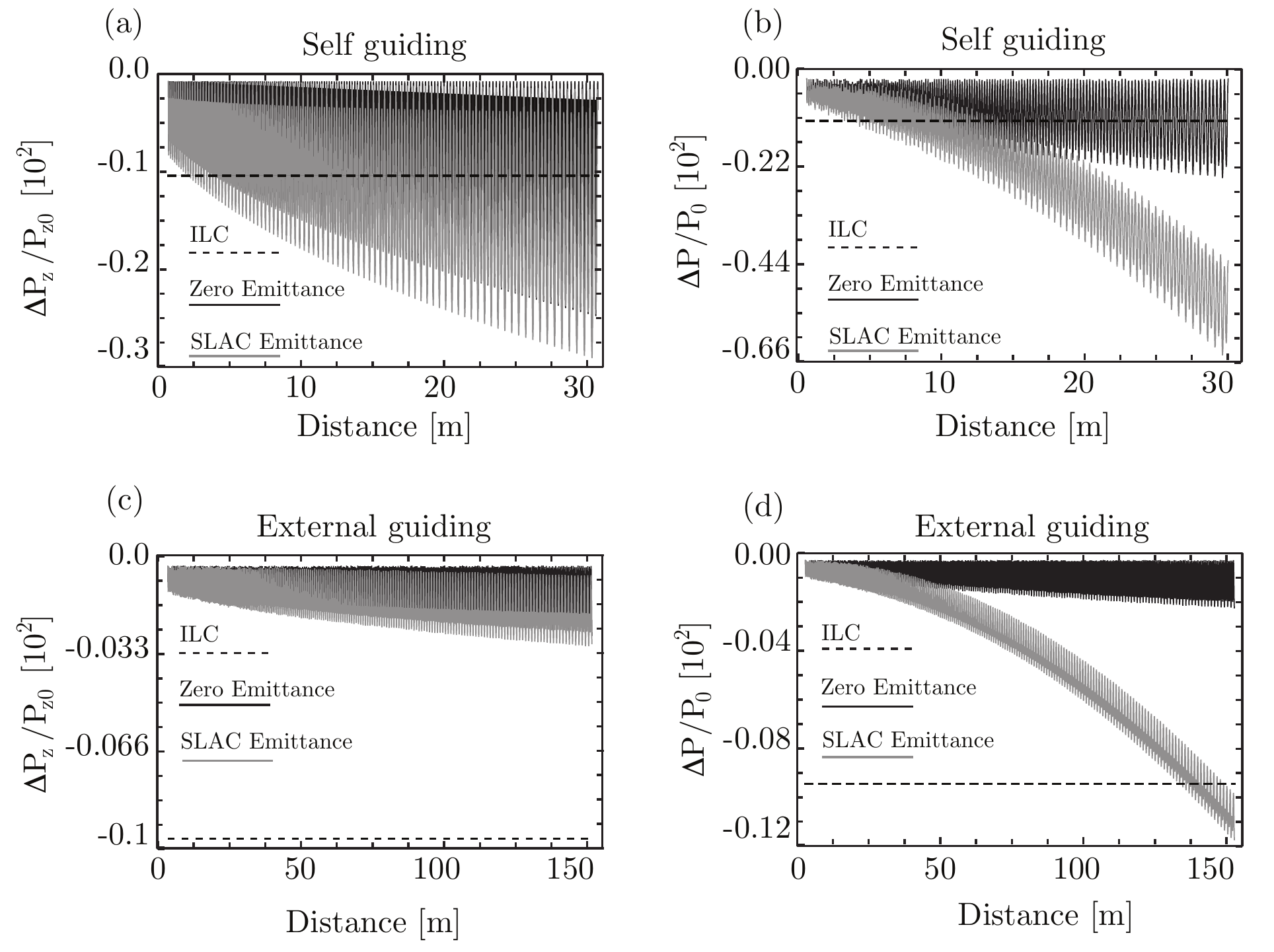}
\caption{\label{fig:SLAC_500GeV} Beam polarization evolution for a 0.5 TeV PBA. (a) Evolution of the longitudinal component of the polarization in a self-guided regime, (b) Evolution of the total polarization in a self-guided propagation regime, (c) Evolution of the longitudinal component of the polarization in an externally guided propagation regime, and (d) Evolution of the total polarization in an externally guided propagation regime. The gray (black) line corresponds to a SLAC-like beam with (without) emittance, and the black dashed line corresponds to the typical depolarization that is expected for the ILC, which includes the depolarization of the beam at the interaction point between the colliding beams.} 
\end{center}
\end{figure}

Using the spin-tracking numerical algorithm, the depolarization associated with realistic electron beams, with finite emittance, can be investigated. Of particular relevance is the assessment of the role of the beam emittance for typical electron beams now available. Fig.~\ref{fig:SLAC_500GeV} shows the evolution of the polarization of a SLAC-like electron beam, initialized with $\hat{\sigma}_r=10~\mu\mathrm{m}$ and $\gamma_0=6\times 10^4$. The beam normalized emittance is $\epsilon_{Nx}=50~\mathrm{mm}\cdot\mathrm{mrad}$ and $\epsilon_{Ny}=20~\mathrm{mm}\cdot\mathrm{mrad}$ in the $x$ and $y$ directions, reaching 0.5 TeV at the end of the acceleration. The plasma parameters were chosen to mimic LWFA stable propagation regimes: for the self-guided case $n_0=1.25\times 10^{16}~\mathrm{cm}^{-3}$, $E_{\mathrm{accel}}=1.65$, and $\alpha=0.5$; for the externally-guided scenario, $n_0=2.4\times10^{15}~\mathrm{cm}^{-3}$, $E_{\mathrm{accel}}=0.7$, and $\alpha=0.35$. The longitudinal depolarization rates are only marginally increased due to the non-zero emittance, which plays a stronger role in the external-guiding scenario. However, the total depolarization rates increase due to a drift of the transverse spin component of single beam electrons, which leads to stronger variations of $\mathbf{P}_{\perp}$ (cf. Eq.~(\ref{eq:p_perp_amp})). The numerical results also show that the effects of the beam emittance on the depolarization can be neglected for the early propagation. The expected total depolarization associated with the ILC, including all the depolarization sources present at a linear collider, is also shown in Fig.~\ref{fig:SLAC_500GeV}. The main depolarization source at the ILC comes from the beam-beam interaction at the interaction point, which provides typical depolarizations on the order of $0.1-0.2\%$. Although this work does not include the depolarization that occurs at the collision of the two colliding beams, Fig.~\ref{fig:SLAC_500GeV} suggests that the main source of depolarization in a plasma based collider may also come from the beam-beam interaction at the interaction point provided that the beam emittance is sufficiently low, so that the beam depolarization is kept below $0.1-0.2\%$ during the acceleration.

3D PIC simulations using the quasi-static PIC code QuickPIC~\cite{bib:huang_jcp_2006} have been performed to examine in detail the spin precession and polarization dynamics in LWFA scenarios. To this purpose, a random sample of the externally injected electrons was tracked using a particle-tracking algorithm~\cite{bib:fonseca_ppcf_2008}, which stored the trajectories and fields associated with the accelerated electron beam.  

The simulations, which worked in the externally guided propagation regime, used the matched parameters from Refs.~\cite{bib:lu_prstab_2007,bib:tzoufras_prl_2008} for a $800~\mathrm{J}$ laser pulse. The laser central wavelength was $\lambda_0=0.8~\mu\mathrm{m}$, normalized peak vector potential $a_0=(e A_0/m_e c^2)=2$, spot size $W_0=140~\mu\mathrm{m}$, and duration $\tau_{\mathrm{FWHM}}=311~\mathrm{fs}$. The transverse plasma profile is given by $n(r)=n_0[1+\Delta n (r/W0)^2]$ for $r<r_c$, falling linearly for $r>r_c$, where $n_0=1.15\times10^{16}~\mathrm{cm}^{-3}$, $\Delta n=0.375$, $r_c=560~\mu\mathrm{m}$, and $r_f=2380~\mu\mathrm{m}$. The size of the simulation box is $2380~\mu\mathrm{m}\times2380~\mu\mathrm{m}\times450~\mu\mathrm{m}$, divided into $128\times128\times 1024$ cells, with $4$ particles per cell. The externally injected electron beam was initialized at the back of the first plasma wave, in the region of maximum accelerating fields, with zero-emmitance, and zero-energy spread. The beam charge was $q=80~\mathrm{pC}$, and $\gamma_0=10^3$. The beam density is transversely Gaussian with width $\sigma_r=23~\mu\mathrm{m}$ and length $\sigma_z=23~\mu\mathrm{m}$. In the simulations, 125 electrons were followed in time, and the particle-tracked data was post-processed according to Eqs.~(\ref{eq:1ststep}) and (\ref{eq:2ndstep}), in order to retrieve the evolution of the beam polarization.        

\begin{figure}[htbp]
\begin{center}
\includegraphics[width=\columnwidth]{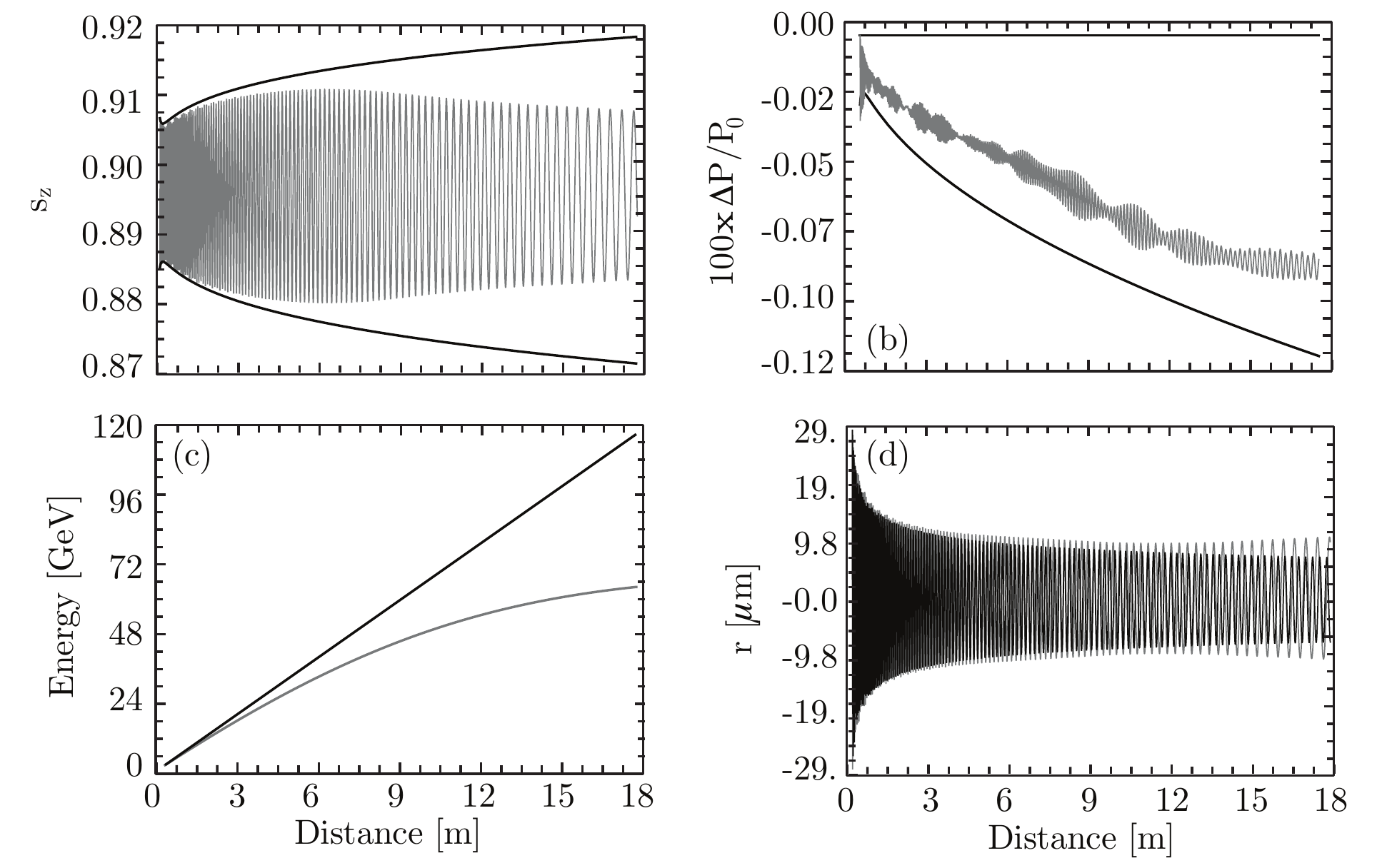}
\caption{\label{fig:polarization} QuickPIC simulation results for a externally guided LWFA, using a $800~\mathrm{J}$ laser pulse to accelerate a zero-emittance electron beam to 70 GeV. The predictions of the model are shown in black, and the results of the simulation shown in gray. (a) shows the evolution of $s_z$ for a randomly chosen beam electron. (b) shows the evolution of the total polarization. The black curves are the analytical results for the maximum and minimum values of $\Delta P/P_0$. (c) shows the evolution of the electron beam energy. (d) shows the radial trajectory of the same electron of (a).} 
\end{center}
\end{figure}

Fig.~\ref{fig:polarization} shows a comparison between the simulation and the analytical model. For the analytical model results, $E_{\mathrm{accel}}=\sqrt{a_0}/2\simeq 0.7$~\cite{bib:lu_prstab_2007} was used, and the coefficient $\alpha=0.35$ was retrieved from the simulations (note that even though in the simulation a parabolic plasma channel is used, the focusing force is still linear in the region of the beam). Excellent agreement is obtained for $s_z$ until $3~\mathrm{m}$ of propagation (Fig.~\ref{fig:polarization}(a)), when $\gamma \simeq 5\times10^4$ ($\simeq 25~\mathrm{GeV}$). Despite the not perfectly uniform acceleration along the electron beam as the acceleration progresses ($E_{\mathrm{accel}}$ drops), very good agreement was also found for the evolution of the polarization (Fig.~\ref{fig:polarization}(b)). We note that the oscillation amplitude of the polarization does not match the analytical model because of the limited number of tracked particles in the 3D simulations, which is limited by computational constraints. However, the trend of the polarization is fully captured. For propagation distances larger than $3~\mathrm{m}$ the model overestimates the spin precession amplitude because of the reduction of the acceleration gradient (due to the laser pump depletion and dephasing), and also because the focusing force changes ($\alpha$ decreases during the simulation).

\begin{figure}[htbp]
\begin{center}
\includegraphics[width=\columnwidth]{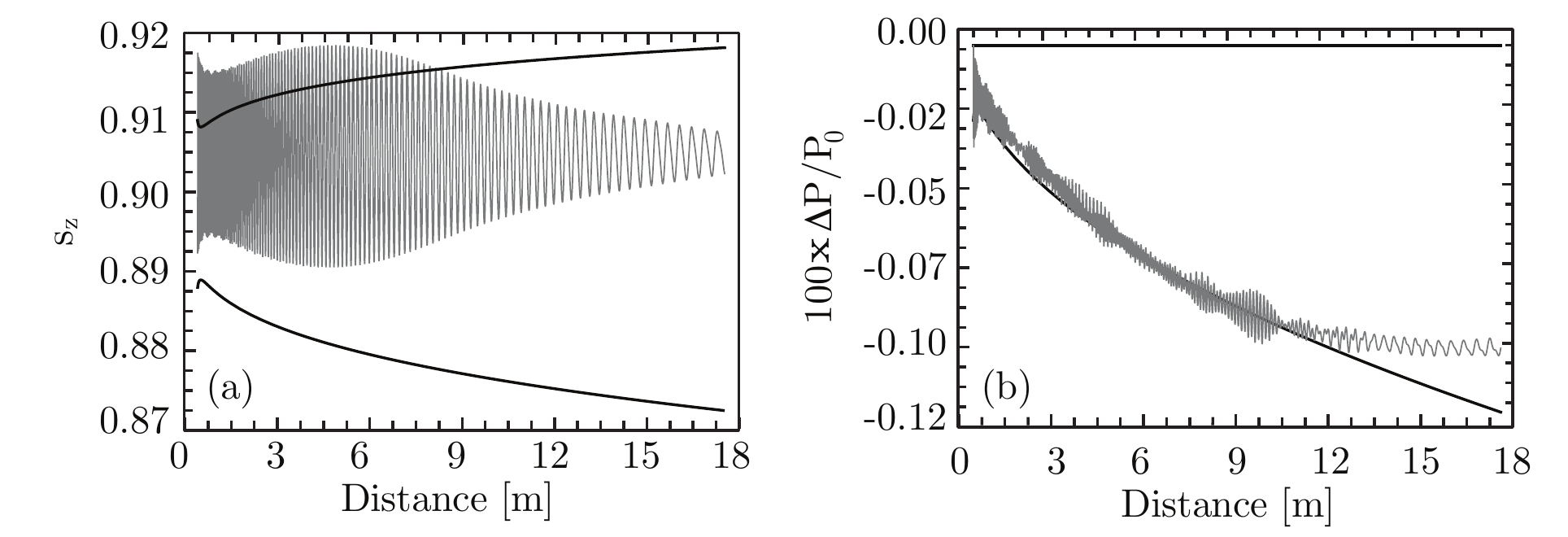}
\caption{\label{fig:polarization_emittance} QuickPIC simulation results for a externally guided LWFA, using a $800~\mathrm{J}$ laser pulse to accelerate an electron beam with normalized emittance $\epsilon_N = 62.5~\mathrm{mm}\cdot{mrad}$ to 70 GeV. The black curves refer to the predictions of the analytical model, and the gray curves represent the simulation results (a) shows the evolution of $s_z$ for a given beam electron. The results of the model assumed no initial transverse electron momentum. (b) shows the evolution of the total polarization. The analytical results illustrate the maximum and minimum values of $\Delta P/P_0$ assuming a zero-emittance electron beam.} 
\end{center}
\end{figure}

In order to identify the role of the beam emittance in more realistic scenarios, QuickPIC~\cite{bib:huang_jcp_2006} simulations using non-zero electron beam emittances were performed, using $\epsilon_N^{\mathrm{sim}} = 62.5~\mathrm{mm}\cdot\mathrm{mrad}$ (typical SLAC beam emittance) and keeping the rest of the laser, plasma, and beam parameters unchanged. For these values, Eq.~(\ref{eq:emittance}) yields $\epsilon_N^{\mathrm{max}} = 167~\mathrm{mm}\cdot\mathrm{mrad}\simeq 3 \epsilon_N^{\mathrm{sim}}$. Fig.~\ref{fig:polarization_emittance}(a) shows the evolution of $s_z$ for a given beam electron. The initial electron velocity shifts the value of $s_z$ from the theoretical prediction because the amplitude of the electron oscillation changes due to the initial electron velocity. The total beam polarization, shown in Fig.~\ref{fig:polarization_emittance}(b), indicates that the depolarization rates are slightly higher in comparison to those of Fig.~\ref{fig:polarization}(b). 


\section{Conclusions}
\label{sec:conclusions}

This Paper describes the key features associated with the acceleration of polarized electron beams in PBAs. It was found that the spin precession dynamics in the PBA is closely related to the betatron oscillations in the ion channel, and that for zero emittance beams, the polarization vector oscillates in the plane defined by the propagation direction and the initial polarization vector direction. Therefore, if the beam is initially polarized in an optimal direction for a certain HEP application, it remains in that direction during the acceleration stage. Furthermore, higher energy matched LWFAs lead to lower depolarizations.

Our model was validated by the numerical resolution of the T-BMT equations, using prescribed electromagnetic plasma wave fields and electron trajectories, and by using the fields and electron trajectories retrieved from 3D PIC simulations in QuickPIC~\cite{bib:huang_jcp_2006}. These results further confirmed the potential of PBA to accelerate polarized electron beams to high energies with final depolarizations which can fulfill the requirements of high energy physics experiments provided that low beam emittances are used.

The analytical work presented in this Paper was obtained in the limit of small emittances such that $\epsilon_{N}[\mathrm{mm} \cdot \mathrm{mrad}] \ll \epsilon_{N}^{\mathrm{mrad}}= 188 \sqrt{\alpha} \left(\hat{\sigma}_r/10~\mu\mathrm{m}\right)^2\left(E_0/10~\mathrm{GeV}\right)^{1/2}\left(n_0/10^{16}~\mathrm{cm}^{-3}\right)^{1/2}$. We note that the beam luminosities are also larger in the conditions where $\epsilon_{N}[\mathrm{mm} \cdot \mathrm{mrad}] \ll \epsilon_{N}^{\mathrm{mrad}}$. Therefore our analytical model describes the spin precession in PBAs in the conditions that are directly relevant for HEP applications. Moreover, numerical modeling showed that larger initial beam emittances lead to larger depolarization rates, specially in the transverse directions. 3D PIC simulations in QuickPIC also indicated that the depolarizing effects due to the beam emittance are balanced by the decrease of the accelerating gradients caused by the laser pump depletion.

In future works it could be important to further assess the effects of the beam emittance, beam energy spread in the beam depolarization, along with the depolarization associated with the beam-beam interaction at the interaction point in plasma based accelerator colliders.

Work partially supported by FCT (Portugal) through grants SFRH/BD/22059/2005, PTDC/FIS/111720/2009, and CERN/FP/116388/2010, EC FP7 through LaserLab-Europe/Laptech; UC Lab Fees Research Award No. 09-LR-05-118764-DOUW, the US DOE under DE-FC02-07ER41500 and DE-FG02-92ER40727, and the NSF under NSF PHY-0904039 and PHY-0936266. Simulations were done on the IST Cluster at IST, on the Jugene supercomputer under a ECFP7 and a DEISA Award, and on Jaguar computer under an INCITE Award. 


\bibliographystyle{apsrev}

\end{document}